\documentstyle[aps]{revtex}
\begin{document}

\noindent

\let\a=\alpha \let\b=\beta  \let\c=\chi
\let\d=\delta  \let\e=\varepsilon
\let\f=\varphi \let\g=\gamma \let\h=\eta
\let\k=\kappa  \let\l=\lambda
\let\m=\mu   \let\n=\nu   \let\o=\omega
\let\p=\pi  \let\ph=\varphi
\let\r=\rho  \let\s=\sigma \let\t=\tau
\let\th=\vartheta
\let\y=\upsilon \let\x=\xi \let\z=\zeta

\let\D=\Delta \let\F=\Phi  \let\G=\Gamma
\let\L=\Lambda \let\Th=\Theta
\let\O=\Omega \let\P=\Pi   \let\Ps=\Psi
\let\Si=\Sigma \let\X=\Xi
\let\Y=\Upsilon

\def\FF{{\cal F}}
\let\io=\infty
\def\pp_F{{\vec p}_F}
\def\V#1{\vec#1}
\def\oo{{\vec \o}} \def\OO{{\vec \O}} \def\uu{{\vec \y}}
\def\xx{{\vec x}} \def\yy{{\vec y}} \def\kk{{\vec k}}
\def\BB{{\cal B}}
\def\nn{\nonumber}
\def\VV{{\cal V}}
\def\E{{\cal E}} \def\ET{{\cal E}^T}
\def\LL{{\cal L}}\def\RR{{\cal R}}\def\SS{{\cal S}}
\def\NN{{\cal N}}
\def\HH{{\cal H}}\def\DD{{\cal D}}\def\GG{{\cal G}}
\def\ie{{\it i.e. }}
%\draft
\title{Incommensurate Charge Density Waves
in the adiabatic Hubbard-Holstein model}
\author{Vieri Mastropietro}
\address{Universit\`a di Tor Vergata,
Roma, Italia}
\date{\today}
\maketitle
\begin{abstract}
The adiabatic, 
Holstein-Hubbard model describes
electrons on a chain with step $a$ 
interacting with
themselves (with coupling $U$)
and with a classical phonon
field $\f_x$ (with coupling $\l$). There is 
Peierls instability 
if the electronic ground state energy $F(\f)$
as a functional of $\f_x$ has a minimum which corresponds to
a periodic function with period ${\pi\over p_F}$,
where $p_F$ is the Fermi momentum. 
We consider
${p_F\over\pi a}$ irrational so that 
the CDW is
{\it incommensurate} with the chain.
We prove in a rigorous way
in the spinless case, when $\l,U$ are small and
${U\over\l}$ large, that 
a)when the electronic interaction is attractive $U<0$
there is no Peierls instability b)when the interaction
is repulsive $U>0$ there is Peierls instability
in the sense that our convergent expansion for $F(\f)$, 
truncated at the second order,
has a minimum which corresponds to an analytical
and ${\pi\over p_F}$ periodic $\f_x$. 
Such a minimum is found solving an infinite set
of coupled self-consistent equations, one for each of the
infinite Fourier modes of $\f_x$. 
\end{abstract}
%\input fiat
%\BOZZA
\section{Introduction}
In 1955 Peierls, in Ref.\cite{[P]}, suggested that in a 
one dimensional metal it is energetically favorable
to develop a periodic distortion of the linear lattice
with period ${p_F\over\pi}$ where $p_F$ 
is the Fermi momentum
of the conduction electrons.
The attempt of the conduction electrons 
to screen the periodic potential generated by
the periodic lattice distortion creates a {\it Charge Density Wave}
(CDW) in the conduction electron density.
If $a$ is the step of the undistorted lattice, then, depending
whether ${p_F a\over\pi}$ is a rational number or not,
the CDW (or the periodic lattice distortion)
can be either {\it commensurate}
or {\it incommensurate} with the non distorted
lattice. While a commensurate CDW has preferred positions
in the lattice, an incommensurate CDW has not
and so it can slide without any change of energy; 
this was considered
by Frolich in Ref.\cite{[F]} a possible mechanism for superconductivity
or a least for an enhancement of conductivity (see Ref.\cite{[LRA]}).

Indeed starting from the 70's both commensurate
or incommensurate CDW with wavevector
$2 p_F$ have indeed been observed in a number of compounds
(see for instance Refs.\cite{[T]} and \cite{[FR]}). A new wind of interest
followed recently the discovery of
high
$T_c$ superconductors showing  
one dimensional
incommensurate CDW, see Refs.\cite{[B]} and \cite{[PCGD]}.

From a theoretical point of view, Peierls instability can be studied
in the
{\it Holstein-Hubbard} model, which 
is the simplest model involving both an electron-phonon
and an electron-electron interaction.
The standard theory of CDW is usually developed within
the adiabatic approximation, in which the phonon field
is treated as a classical field and the model becomes variational.
We will consider the spinless case,
so that the {\it spinless adiabatic Holstein-Hubbard} hamiltonian is
\begin{eqnarray}
&&H=H_0+H_p+\l P+U V=\label{1}\\
&&\sum_{x,y\in\L}
(t_{xy}-\mu\d_{x,y})\,\psi^+_{x}\psi^-_{y}+{1\over2}\sum_{x\in\L}
\f^2_x-\l \sum_{x\in\L} \f_x\psi^+_{x}\psi^-_{x}+U \sum_{x \in\L}
[\psi^+_{x}\psi^-_{x}-{1\over 2}][\psi^+_{x+1}\psi^-_{x+1}-{1\over 2}],\nn
\end{eqnarray}
where
$\L=1,...,L-1$, $t_{x,y}=\d_{x,y}-{1\over
2}(\d_{x,y+1}+\d_{x,y-1})$,
$\psi_{x}^\pm$ are fermionic creation or annihilation
fields with periodic boundary conditions, $\mu$ is
the chemical
potential $\mu=1-\cos p_F$.
In (\ref{1}) $H_0$ is the fermionic kinetic energy,
$H_p$ is the phonon kinetic energy, $\l P$
is the electron-phonon interaction and $U V$
is the electron-electron interaction; to describe the Coulomb
repulsion one needs $U>0$ but it is not irrealistic 
to consider also $U<0$ (in this case $U V$
is an effective interaction 
taking into account phonon-mediated processes).
If $U=0$ the above model is called {\it Holstein model}.
$p_F$ is the Fermi momentum of the non interacting $U=0$ model;
the Fermi momentum in the $U\not=0$ case is in general different, but we fix
it to $p_F$ by adding a term $\nu N$ to the hamiltonian, where
$N=\sum_{x\in\L}\psi^+_x\psi^-_x$ is the total particle number operator and $\nu$ is a suitable counterterm.
The proof of Peierls instability
consists, within this model, in the proof that
the ground state energy 
$F(\f)=\sum_{x\in\L}{\f_x^2\over 2}+E_0(\f)$ is minimized
by $\f_x=\bar\f(2 p_F x)$ 
where $\bar\f(t)$ is a $2\pi$-periodic function. 
The existence of a global minimum of the form 
$\bar\f(2 p_F x)$  was proved in 
the half filled band case $p_F=\pi/2$
in the Holstein model 
and in the spinning Holstein-Hubbard model (see Refs.\cite{[KL]}
and \cite{[LN]}).
Local minima of the form $\bar\f(2 p_F x)$,
for any $p_F=\pi{P\over Q}$ with $P,Q$ relatively prime, 
were found in Ref.\cite{[BGM2]},
for $|\l|\le O(\log Q^{-1})$.
We are interested here in the {\it incommensurate} case in which ${p_F\over\pi}$ is an irrational number.
At finite $L$, 
it is not possible to fix ${p_F\over\pi}$ directly to an irrational
number, as in this way $\bar\f(2 p_F x)$
cannot verify periodic boundary conditions.
We look however to a sequence 
of $L_i$, $n_i$ such that $\lim_{i\to\io} L_i=\io$
and 
$\lim_{i\to\io} p_{F,i}=p_F$
and
${p_F\over\pi}$ is irrational, where $p_{F,i}={2\pi n_i\over L_i}$. An {\it incommensurate} phonon
field has the form
\begin{equation}
\f_x=\lim_{i\to\io}\sum_{n=-[L_i/2]}^{[(L_i-1)/2]}\hat\f_n e^{i2 p_{F,i} n x}.\label{1.3}
\end{equation}
We require moreover
that
\begin{equation}
\|2n p_{F,i} \|_{\bf T^1} \ge C_0 |n|^{-\t} \; , \quad
0\neq n \in Z, \;\qquad |n|\le {L_i\over 2},\label{11}
\end{equation}
where $||k||_{\bf T^1}=\min_{n\in Z}|k-2\pi n|$.
(\ref{11}) means that 
$p_F/\pi$ is a {\it Diophantine number} (see for instance Ref.\cite{[G]}) and the proof of the existence
of $p_{F,i}$ verifying (\ref{11}) can be found in Ref.\cite{[BGM1]}. 
Such assumption is not really restrictive, as
if $\t>1$ the complementary set of such points has measure $0$.
A Diophantine condition like (\ref{11}) (in the infinite volume limit)
appears in classical mechanics, for instance in the KAM theorem, see
Refs. \cite{[A]} and \cite{[Mo]}, 
and it is useful for handling with the so called
{\it small divisor problem}; a similar problem appears also here.
That a sort of extension of KAM-techniques to quantum systems
is necessary to prove Peierls instability in the incommensurate case
was pointed out for the first time by Aubry in Ref.\cite{[AAR]} by analogy
with the {\it Frenkel-Kontorova} models (see also Ref.\cite{[AL]}).
We have to specify the space of functions on which $F(\f)$ is defined
as a variational form.
We say that
$F(\f)$ is a functional $F:\O\to R$ where
{\it 
$\O$ is the set of functions $\f_x$ 
of the form (\ref{1.3}) 
with zero average $\hat\f_0=0$ and  
$\hat \f_n=\hat \f_{-n}=\hat \f_n^*$.
Moreover, if $\k, F_0, F_1>0$ are constants and $\s=\l\hat\f_1$
then $|\s|\le F_0$ and, for $|n|>1$
\begin{equation}
|\l\hat\f_n| \le F_1 |\s| e^{-\k|n|}.
\label{1.12} 
\end{equation}}
\*
The condition (\ref{1.12})
ensures also that the 
$2 p_F$ harmonic is present (
if $\hat\f_1=0$ then $\f_x$
is a constant). 
If $\f$ is an extremal point of $F(\f)$, it must satisfy the
condition $\hat\f_0=\l\r$ where $\r$ is the fermionic density.
On the other hand, 
we can always include $\hat\f_0$ in the chemical potential $\mu$
and then we can
restrict our search of
local minima of the ground state energy $F(\f)$ to fields $\f$ with zero mean.
Note that $F:\O\to R$ is indeed a function of the Fourier
coefficients
$\hat\f_n$; this means 
that, at finite $L$, can be considered
not a functional but
a $L$-dimensional function 
and only at the end we will take the $L\to\io$ limit. 

We say that  {\it there  is Peierls instability 
if the variational form $F:\O\to R$ has a minimum
$\f_x\in\O$} .  
We will show in \S 2 
that if $\f_x\in\O$ is a local minimum then
\begin{equation}
\hat\f_n=
\l\hat\r_n(\f)\quad n\not=0,\quad n=-[L/2],...,[(L-1)/2],\label{2.5}
\end{equation}
and
$M_{nm}=\d_{nm}-\l {\partial\over\partial\hat\f_n}\hat\r_m(\f)$
positive definite,
where $\hat\r_n(\f)={\partial E_0(\f)\over\partial\hat\f_n}$.
Peierls instability can be proved by solving
the infinite (as $L\to\io$) set of coupled equations (\ref{2.5}).
There are then two main steps to be performed; the first is to compute
$\hat\r_n$ by an expansion, as there is no hope to compute 
it in a simple exact form,
and the second is to solve the system $\hat\f_n=
\l\hat\r_n(\f)$. 
In \S 2 an expansion for $\hat\r_n(\f)$ is found, which
is {\it convergent} for any $\f\in\O$, for $\l,U$ small enough.
The proof of convergence
is based on a sort of generalization of KAM theorem to quantum system;
in \S 2 we review the main ideas referring to Ref.\cite{[M1]} for the mathematical proofs.
The result is that we can write
$\hat\r_n=\sum_{k=0}^\io\hat\r_n^{(k)}$ and, {\it if $p_{F,i}$ verifies
(\ref{11}) and $\f_x\in\O$, then for $|\l|,|U|\le \e$ one has
$|\hat\r^{(k)}_n|\le f(n,\l,U) C^k\e^k$, where $C$ is a constant and $f(n,\l,U)$
is a  proper function}. Of course the exact form of $f(n,\l,U)$ is important, and 
it will be
specified in \S 2.E. The 
proof of the convergence (for $\e$ small enough) of the expansion
for the density is based on Renormalization Group
methods; it is important however to stress that, while the usual
RG methods are only approximative, our results are {\it mathematically
exact}; we refer for an introduction to rigorous RG
for fermions to Ref.\cite{[GM]}.

The second step (see \S 3)  consists in solving
(\ref{2.5}). This is quite a difficult task,
and we consider the simpler problem obtained by keeping 
only the first two terms of the expansion for $\hat\r_n$.
This means that we study
\begin{equation}
\hat\f_n=
\l\hat\r^{(0)}_n(\f)+\l\hat\r_n^{(1)}(\f),\qquad n=-[L/2],...,[(L-1)/2].\label{2.5a}
\end{equation}
so neglecting $O(\e^2)$ terms in the r.h.s. of (\ref{2.5}).
The convergence of the expansion for $\hat\r_n$ makes
this approximation quite reasonable.
Note also that $\hat\r_n^{(0)}$ and  
$\hat\r_n^{(1)}$ are rather complex function of $\l,U$ (the expansion is not
a power series in $\l, U$). Our main result is that  {\it if ${|\l|\over |U|}$ is small enough, in the $L\to\io$ limit

a)in the attractive
$U<0$ case there are no solutions $\f\in\O$ of ({\ref{2.5a}) (and of (\ref{2.5}) as well);

b)in the repulsive case $U>0$ there is a $\f_x\in\O$
solving (\ref{2.5a}) such that 
\begin{equation}\label{jk}
\l\hat\f_1\equiv\s=A[{\l^2\over a\h}]^{1\over \h}
[1+O({\l^2\over U})]^{1\over \h};\qquad
|\hat\f_n|\le e^{-{|\log|\l||\over 10}|n|}|\s|\qquad |n|\not=1.
\end{equation}
with $\h=\b_1 U+O(U^2)$ is a critical index and $a,A$ positive constants.}}
This means that, if the electron-electron interaction
is larger than the electron-phonon interaction, 
there is a dramatic dependence on the repulsive or attractive nature
of the electron-electron interaction.
In the attractive case
there is no Peierls instability  as there are no functions analytical and
${\pi\over p_F}$ periodic minimizing the ground state energy.
In the repulsive case there is instead
Peierls instability, in the sense that the energy,
keeping the first two non trivial terms of its convergent expansion,
has a minimum which corresponds to an incommensurate CDW. 
It is very reasonable that higher order terms do not change
this result (see the considerations in \S 3.B). 

Our results improve preceding works on the subject 
in which a)the interaction among electrons was neglected; b)the analysis
was restricted to the {\it first order} 
\begin{equation}
\hat\f_n=\l\hat\r^{(0)}_n(\f),\label{2.5c}
\end{equation} 
where, if $U=0$,  $\l\hat\r^{(0)}_n(\f)\simeq -a \l^2\hat\f_1\log|\l\hat\f_1|$
if $n=\pm 1$ and zero otherwise.
There is a major difference between (\ref{2.5c})
and (\ref{2.5a}); while (\ref{2.5c}) admits trivially a solution $\l\hat\f_n=\s\d_{n,\pm 1}$
and $\s$ is obtained by a BCS like equation, a solution of the form 
$\l\hat\f_n=\s\d_{n,\pm 1}$ {\it does not solve} (\ref{2.5a}) or (\ref{2.5}).
While a solution of (\ref{2.5c}) if $U=0$ 
(when the model reduces to the Holstein model)
is trivial to find, we are not
able to find a solution of (\ref{2.5a}) $\in\O$ if $U=0$,
as the method we use to find a solution of (\ref{2.5a})
when $U>0$ and ${U\over|\l|}$ is large fails in that case.
We have then no evidence of Peierls instability in the Holstein model
and it is unclear if this is merely a technical problem or 
if there is $U_c>0$ such that
Peierls instability holds only for $U>U_c$ in the Holstein-Hubbard model.
\vskip.5cm
The paper is organized in the following way.
In \S II.A we describe an expansion for the ground state energy
of the Hubbard-Holstein model (\ref{1})
for $\f\in\O$. 
In order to prove
the convergence (for small $\l,U$) of the expansion 
for small $\l,U$, one has to solve a small
divisor problem, and this is discussed in \S II.B.
In \S II.C it is studied the RG flow and in \S II.D
we prove that the renormalization of the Fermi momentum
if $\l$-independent. From the ground state energy
expansion, is it easy to derive
an expansion for the density, and this is done in \S II.E.
In \S III we prove that (\ref{2.5}) has no solution $\f\in\O$
in the attractive case, if ${|\l|\over |U|}$
is small enough, while in the repulsive case
we find a solution $\f\in\O$ of (\ref{2.5a}) by a contraction method.
Finally in \S IV we discuss 
some open problems, in particular for the $U=0$ case.

\section{Renormalization Group analysis}
\subsection{Grassman integrals}

It is well known that $E_0(\f)$
can be written as a {\it Grassman integral} (
we use the same symbol $\psi$ for
field and Grassman variables with a traditional abuse of notation)
\begin{equation}\label{la}
E_0(\f)=-\lim_{\b\to\io}{1\over L\b}\log \int P(d\psi) e^{-UV-\l P-\nu N},
\end{equation}
where
\begin{equation}\label{la1}
V=\int_{-{\b\over 2}}^{\b\over 2} d x_{0}
\sum_{x\in\L}
[\psi^+_{\xx}\psi^-_{\xx}-{1\over 2}][\psi^+_{\xx+1}\psi^-_{\xx+1}-{1\over
2}],\quad P=-\int_{-{\b\over 2}}^{\b\over 2} d x_{0} \sum_{x\in\L}\f(x)
\psi^+_{\xx}\psi^-_{\xx},\quad N=
\int_{-{\b\over 2}}^{\b\over 2} d x_{0} \sum_{x\in\L}
\psi^+_{\xx}\psi^-_{\xx},
\end{equation}
and $\xx=(x_0,x)$ and $\xx+1=(x_0,x+1)$. $P(d\psi)$ is a
{\it Grassmanian integration} defined on monomials by the anticommutative
Wick rule with propagator
\begin{equation}
g(\xx;\yy)={1\over\b L}\sum_\kk{e^{-i\kk(\xx-\yy)}\over -i k_0-\cos k+\cos
p_F},
\end{equation}
where $\kk=(k_0,k)$.
(\ref{la}) has a well defined $L,\b\to\io$
limit only if the counterterm $\nu$ is chosen
in a suitable way as a function of the parameters
appearing in the Hamiltonian so that the Fermi momentum
is just $p_F$.  
In order to find
the minima of $F(\f)$ we have to differentiate with
respect to $\hat\f_n$, so one has in principle
to take into account the possible dependence
of $\nu$ from $\hat\f_n$, which is in general
very complicated.
However we will show in \S 2.D that it is possible
to choose $\nu$ as {\it independent} from $\l$ and so from $\hat\f_n$. 
This is due to the fact that the chemical potential can
be moved inside the gap opened by $\f_x$ without
affecting any physical property, and we can use
this freedom to fix $\nu$ as independent of $\hat\f_n$.
It follows that a necessary condition for $\f_x\in\O$ 
to be a local minimum for $F(\f)$ is that
it verifies
$\f_x=\l\r_x$ where
$\r_x=\lim_{\b\to\io,\t\to 0}{1\over L}S^{L,\b}(x,\t;x,0)$
and $S^{L,\b}(x,\t;x,0)$ is the Schwinger function defined by,
if $\phi^\pm_\xx$ are Grassman variables
and writing $\int d\xx=\int_{-\b\over 2}^{\b\over 2}\sum_{x\in\L}$,
\begin{equation}
S^{L,\b}(\xx;\yy)={\partial^2\over\phi^+_\xx\partial\phi^-_\yy}
\log\int P(d\psi)e^{-\VV(\psi)-\int
d\xx[\phi^+_\xx\psi_\xx^-+\phi^-_\xx\psi_\xx^+]}|_{\phi=0},
\end{equation}
and $\VV=U V+ \l P+\nu N$. (\ref{2.5}) is obtained by
the Fourier transform of $\f_x=\l\r_x$
defining $\r_x=\sum_{n=-{[L/2]}}^{[(L-1)/2]}e^{2 i n p_F x}\hat\r_n$.

The above Grassman integrals can be evaluated by Renormalization Group 
methods; we refer to Ref.\cite{[GM]} for an introduction 
to the formalism we are using and to Ref.\cite{[M1]} for the mathematical
proofs of the convergence of the expansion we are describing.
We
start by evaluating the partition function
$\int P(d\psi) e^{-\VV(\psi)}$.
It is convenient to decompose the Grassman
integration $P(d\psi)$ into a product 
of independent integrations.
Let be $|\kk|=\sqrt{k_0^2+||k||_T^2}$.
We write 
\begin{equation}
g(\kk)=f_1(\kk) g(\kk)+(1-f_1(\kk)) g(\kk)=
g^{(u.v.)}(\kk)+g^{(i.r.)}(\kk),\label{3.3}
\end{equation}
where $f_1(\kk)=1-\chi(k-p_F,k_0)-\chi(k+p_F,k_0)$ and
$\chi(k',k)$ is a $C^\io$ function
with compact support such that it is $1$ 
for $|\kk'|\le{a_0\over\g}$ and $0$ for 
$|\kk'|>a_0$, where $\g>1$ and $\g,a_0$ are chosen
so that $\chi(k\pm p_F,k_0)$ are non vanishing only
in two non overlapping regions around $\pm p_F$.
We write $k=k'+\o p_F$, $\o=\pm 1$ and
\begin{equation}
g^{(i.r.)}(\kk)=\sum_{\o=\pm 1}\sum_{h=-\io}^0 f_h(\kk')g(\kk)\equiv
\sum_{\o=\pm 1}\sum_{h=-\io}^0 g^{(h)}(\kk),
\end{equation}
where $f_h(\kk')=\chi(\g^{-h}\kk')-\chi(\g^{-h+1}\kk')$
has support $O(\g^h)$ around $\o p_F$. 
The integration of $\psi^{(1)}$, the {\it ultraviolet} integration,
gives 
\begin{equation}
e^{-\VV^{(0)}(\psi^{(\le 0)})}=
\int P(d\psi^{(1)}) e^{-\VV(\psi^{(1)}+\psi^{(\le 0)})},
\end{equation}
where $\psi^{\le 0}=\sum_{k=-\io}^0\psi^{(k)}$ and 
(denoting ${1\over\b L}\sum_{\kk}$
simply by by $\int d\kk$)
\begin{equation}
\VV^{(0)}(\psi^{(\le 0)})=
\sum_{n=1}^\io\sum_{m=0}^\io \int d\kk_1...d\kk_{2n}
\psi^{(\le 0)\s_1}_{\kk_1}...
\psi^{(\le
0)\s_{2n}}_{\kk_{2n}}W_{2n,m}^{0}(\kk_1,...,\kk_{2n})\d(\sum_{i=1}^{2n}\s_i\kk_i+2m \pp_F),
\label{v}
\end{equation}
where $\pp_F=(p_F,0)$, $\sigma_i=\pm$ and the kernels $W_{n,m}^{0}(\kk_1,...,\kk_n;z)$ are $C^\io$ bounded functions such
that $W^0_{n,m}=W^0_{n,-m}$ and $|W^0_{n,m}|\le C^n z^{\max(2,n/2-1)}$ if
$z=Max(|\l|,|U|,|\nu|)$. By an explicit computation
it follows that $W^0_{4,0}=U+O(U^2)$, $W^0_{4,m}=O(U\s)$ for
$m\not=0$ and $W^0_{2,m}=\s+O(\s U)$ for $m\not=0$.
$\VV^{(0)}$ is called {\it effective potential at scale $0$};
note that it contains non local interactions between an arbitrary
number of fermions.

The study of the {\it infrared} integration is much more involved.
Following Wilsonian Renormalization group methods, we
have to identify the relevant, irrelevant and marginal interactions;
this is done by a power counting argument and it turns out, as it is
standard in fermionic one dimensional systems, that the interactions
quadratic in the fields are relevant, the quartic are marginal
and the other processes are irrelevant. However
there are in this model $L$ many different terms
bilinear or quartic in the fields, depending on the value of $m$
in (\ref{v}), and so it seems that there are $L$ different
non irrelevant interactions to be taken into account, which seems
an hopeless task as we are interested in the
$L\to\io$ limit; this problem will be solved
by a {\it improved power counting} in which the Diophantine
condition plays a crucial role. Note finally that the quadratic interaction
have a non trivial flow, so it is necessary to change the fermionic
integration at each step; in other words the model has an {\it anomalous
behaviour}, due to the fact that the model is close to a Luttinger liquid.

The integration is performed iteratively, setting $Z_0=1$, $\s_0=\s$,
in the following way: once that the fields $\psi^{(0)},...,\psi^{(h+1)}$ have
been integrated we have
\begin{equation}
\int P_{Z_h}(d\psi^{(\le h)}) \, 
e^{-\VV^{(h)}(\sqrt{Z_h}\psi^{(\le h)})}.\label{3.22}
\end{equation}
Then, putting $C_h(\kk')^{-1}=\sum_{j=-\io}^h f_j(\kk')$ and
$\a(k')=(\cos k'-1)\cos p_F$, $v_0=\sin p_F$:

\begin{eqnarray}\label{3.10b}
&&P_{Z_{h}}(d\psi^{(\le h)}) = 
\prod_{\kk'}\prod_{\o=\pm1} d\psi^{(\le h)+}_{\kk'+\o\pp_F,\o}
d\psi^{(\le h)-}_{\kk'+\o\pp_F,\o}\\
&&\exp \Big\{ -\sum_{\o=\pm1} \int d\kk'
C_h(\kk') Z_{h}
\Big[\Big( -ik_0-\a(k') +\o v_0\sin k' \Big)
\psi^{(\le h)+}_{\kk'+\o\pp_F,\o}
\psi^{(\le h)-}_{\kk'+\o\pp_F,\o}
- \sigma_{h}(\kk') \, \psi^{(\le h)+}_{\kk'+\o\pp_F,\o}
\psi^{(\le h)-}_{\kk'-\o\pp_F,-\o} \Big]
\Big\}.\nn
\end{eqnarray}
Note that after $|h|$ steps the integration is different 
to the initial one; there is a wave function renormalization
$Z_h$ and a mass term $\s_h$.
Moreover the effective potential at scale $h$ has the form
\begin{equation}\label{ty}
\VV^{(h)}(\psi^{(\le h)})=
\sum_{n=0}^\io\sum_{m=0}^\io
\int d\kk'_1...d\kk'_{2n}\prod_{i=1}^n \psi^{\sigma_i
(\le h)}_{\kk'_i+\o_i \pp_F,\o_i}
\d(\sum_{i=1}^{2n} \sigma_i(\kk'_i+\o_i
\pp_F)+2m\pp_F)
W_{2n,m}^{h}(\kk'_1+\oo_1 \pp_F,..;\{\o\}).\nn
\end{equation}
In order to integrate $\psi^{(h)}$ we write
$\VV^{(h)}$ as $\LL \VV^{(h)}+\RR \VV^{(h)}$, with $\RR=1-\LL$.
The $\LL$ operation is defined to extract the non irrelevant terms
in $\VV^{(h)}$; it is easy to check from a power counting argument 
that the terms in $\VV^{(h)}$ involving six
or more fields are irrelevant, thus $\LL=0$ on such terms.
Moreover, we will define $\LL=0$
on the addenda in (\ref{ty}) 
{\it not} verifying the condition
\begin{equation}
\sum_{i=1}^{2n}\s_i\o_i p_F+2 m p_F=0 \quad {\rm mod.}\; 2\pi,\label{f1}
\end{equation}
which means that we are considering irrelevant the terms
such that the sum of momenta measured from the Fermi surface
is not vanishing (but it can be arbitrary small, due to the
irrationality of ${p_F\over\pi}$). 
At the moment this definition of $\LL$ is completely arbitrary;
it will be clear in the next section where we will show
that the terms non verifying (\ref{f1}) are indeed irrelevant (here
is where the Diophantine condition plays a role). 

In conclusion
the definition of $\LL$ is the following:

1) If $2n>4$ then
$$\LL W_{2n,m}^{h}(\kk_1,...)=0$$
2) If $2n=4$ then
\begin{equation}
\LL W_{4,m}^{h}(\kk_1,...)
=\d_{m,0}\d_{\sum_{i=1}^{4}\s_i\oo_i,0} 
W_{4,m}^{h}(\o_1  \pp_F,...,
\o_4 \pp_F)\label{loc1}
\end{equation}

3)If $2n=2$, $\o_1=\o_2$ then
\begin{eqnarray}
&&\LL\{ 
W_{2,m}^{h}(\kk'_1+\o_1 \pp_F,
\kk'_2+\o_2 \pp_F)=\d_{m ,0}
[W_{2,m}^{h}(\o_1 \pp_F,\o_2 \pp_F)\nn\\
&&+\o_1E(k'+\o_1 p_F)\partial_{k}
W_{2,m}^{h}(\o_1 \pp_F,\o_2 \pp_F)+
k^0 \partial_{k_0}W_{2,m}^{h}(\o_1 \pp_F,\o_2 \pp_F)],\label{loc2}
\end{eqnarray}
where $E(k'+\o p_F)=v_0\o\sin k'+(1-\cos k')\cos p_F$ and the symbol
$\partial_k,\partial_{k_0}$ means discrete derivatives. 

4)If $2n=2$, $\o_1=-\o_2$ then
\begin{equation}\label{loc3}
\LL
W_{2,m}^{h}(\kk'_1+\o_1 \pp_F,
\kk'_2+\o_2 \pp_F)
=\d_{m ,\o_2}W_{2,m}^{h}(\o_1 \pp_F,\o_2 \pp_F).
\end{equation}

The Kronecker deltas in the r.h.s. of (\ref{loc1}),
(\ref{loc2})(\ref{loc3}) ensure that $\LL=0$ if (\ref{f1})
is not verified.

We find
\begin{equation}\label{3.11a}
\LL\VV^{(h)}(\psi)=\g^h n_h+F_\nu^{(\le h)}+
s_h F_\s^{(\le h)}+z_h F_\z^{(\le h)}+a_h
F_\a^{(\le h)}+u_h F_U^{(\le h)},
\end{equation}
where
\begin{eqnarray}\label{ccc}
&&F_\s^{(\le h)}=\sum_{\o=\pm 1} \int d\kk'
\psi^{(\le h)+}_{\kk'+\o \pp_F,\o}
\psi^{(\le h)-}_{\kk'-\o \pp_F,-\o},\nn\\
&&F_i^{(\le h)}=\sum_{\o=\pm 1}\int d\kk' f_i(\kk')
\psi^{(\le h)+}_{\kk'+\o \pp_F,\o}
\psi^{(\le h)-}_{\kk'+\o \pp_F,\o},\\
&&
F_U^{(\le h)}=\int [\prod_{i=1}^4 d\kk'_i]\d(\sum_{i=1}^4\s_i\kk_i)
\psi^{(\le h)+}_{\kk'_1+\pp_F,1}
\psi^{(\le h)-}_{\kk'_2+\pp_F,1}
\psi^{(\le 0)+}_{\kk'_3-\pp_F,-1}
\psi^{(\le 0)-}_{\kk'_4-\pp_F,-1}\d(\sum_{i=1}^4\s_i\kk_i),\nn
\end{eqnarray}
where $i=\nu,\z,\a$ and $f_\nu=1$, $f_\z=-i k_0$
and $f_\a=E(k'+\o p_F)$; moreover
$u_0=U(\hat
v(0)-\hat v(2 p_F))+O(U^2)$,
$s_0=O(U \l)$, 
$a_0,z_0=O(U^2)$,
$n_0=\nu+O(U)$. 
Note that in $\LL V^{(h)}$ there are terms
renormalizing mass and the wave function renormalization and
it is convenient to include them in the fermionic
integration writing
\begin{equation}
\int P_{Z_h}(d\psi^{(\le h)}) \, e^{-\VV^{(h)}(\sqrt{Z_h}\psi^{(\le
h)})}=
\int \tilde P_{Z_{h-1}}(d\psi^{(\le h)}) \, e^{-\tilde\VV^{(h)}(\sqrt{Z_h}\psi^{(\le h)})}
\label{3.22a}
\end{equation}
where
$\tilde  P_{Z_{h-1}}(d\psi^{(\le h)})$ is defined as 
$P_{Z_{h}}(d\psi^{(\le h)})$ eq(\ref{3.10b}) with $Z_{h-1}$ and $\s_{h-1}$
replacing $Z_h,\s_h$, with
\begin{equation}
Z_{h-1}(\kk')=Z_h(1+C_h^{-1}(\kk')z_h);\qquad 
Z_{h-1}(\kk')
\s_{h-1}(\kk')=Z_h(
\s_h(\kk')+C_h^{-1}(\kk') s_h.
\end{equation}
Moreover 
$\tilde
\VV^{(h)}=\LL\tilde\VV^{(h)}+(1-\LL)\VV^{(h)}$ and
\begin{equation}\label{3.11b}
\LL\tilde\VV^{(h)}=\g^h n_h F_\nu^{(\le h)}+(a_h-z_h)
F_\a^{(\le h)}+u_h F_U^{(\le h)}
\end{equation}

The r.h.s of (\ref{3.22a}) can be written as
\begin{equation}\label{ml1}
\int P_{Z_{h-1}}(d\psi^{(\le h-1)}) \int \tilde
P_{Z_{h-1}}(d\psi^{(h)}) \, e^{-\tilde \VV^{(h)}(\sqrt{Z_h}\psi^{(\le h)})}
\end{equation}
where $ P_{Z_{h-1}}(d\psi^{(\le h-1)})$ and $\tilde P_{Z_{h-1}}(d\psi^{(h)})$ are given
by (\ref{3.10b}) with $Z_{h-1}$ replaced by $Z_{h-1}(0)\equiv Z_{h-1}$
and
$C_h(\kk')$ replaced with
$C_{h-1}(\kk')$
and $\tilde f_h^{-1}(\kk')$ respectively, if
\begin{equation}
\tilde f_h(\kk')=Z_{h-1}[{C_h^{-1}(\kk')\over Z_{h-1}(\kk')}-
{C_{h-1}^{-1}(\kk')\over Z_{h-1}}]
\end{equation}
and $\psi^{(\le h)}$ replaced with
$\psi^{(\le h-1)}$ and $\psi^{(h)}$ respectively. Note that $\tilde f_h(\kk')$
is a compact support function, with support of width $O(\g^h)$
and far $O(\g^h)$ from the "singularity" \ie $\o p_F$.
The Grassmanian integration $\tilde P_{Z_{h-1}}(d\psi^{(h)})$
has propagator 
$$g^{h}_{\o,\o'}(\xx-\yy)=\int \tilde P_{Z_{h-1}}(d\psi^{(h)})
\psi^-_{\o,\xx}\psi^+_{\o',\yy}$$ 
given by
\begin{equation}
{1\over Z_{h-1}}\int d\kk' e^{-i\kk'(\xx-\yy)}{1\over A_{h-1}(k')}\tilde f_h(\kk')
\left(\begin{array}{c c}
-ik_0-\a(k')-v_0\sin k'&
\sigma_{h-1}(k')\\
\sigma_{h-1}(k')&
-ik_0-\a(k')+v_0\sin k'
\end{array}\right),\label{pr11}
\end{equation}
where $ A_{h}(\kk')=
[ -ik_0-\a(k') ]^2 - (v_0\sin k')^2
- [\sigma_{h-1}(\kk')]^2$.
It is convenient to write decompose the propagator as
\begin{equation}\label{klo1}
g^{(h)}_{\o,\o}(\xx-\yy)=
g^{(h)}_{L,\o}(\xx-\yy)+C_2^{(h)}(\xx-\yy),
\end{equation}
where
\begin{equation}\label{klo2}
g^{(h)}_{L,\o}(\xx-\yy)={1\over L\b}\sum_{\kk'} 
{e^{-i\kk'(\xx-\yy)}\over -i k_0+\o v_0\sin k'+\a(k')}\tilde f_h(\kk')
\end{equation}
and, for any integer $N>1$ 
\begin{equation}\label{klo3}
|g^{(h)}_{L,\o}(\xx-\yy)|\le 
{\g^h C_N\over 1+(\g^h|\xx-\yy|)^N},\qquad
|C_2^{(h)}(\xx-\yy)|\le |{\sigma_h\over \g^h}|^2
{\g^h C_N\over 1+(\g^h|\xx-\yy|)^N}.
\end{equation}
Moreover
\begin{equation}\label{klo4}
|g^{(h)}_{\o,-\o}(\xx-\yy)|\le |{\sigma_h\over \g^h}|
{\g^h C_N\over 1+(\g^h |\xx-\yy)|)^N}
\end{equation}

Finally we {\it rescale} the fields so that
\begin{equation}
\int P_{Z_{h-1}}(d\psi^{(\le h-1)}) \int \tilde
P_{Z_{h-1}}(d\psi^{(h)})
\, e^{-\hat\VV^{(h)}
(\sqrt{Z_{h-1}}\psi^{(\le h)})}
\end{equation}
where
\begin{equation}\label{vbv}
\LL\hat\VV^{(h)}(\psi)=
\g^h\nu_h F_\nu^{(\le h)}+\d_h
F_\a^{(\le h)}+U_h F_U^{(\le h)},
\end{equation}
and by definition
\begin{equation}
\nu_h={Z_h\over Z_{h-1}}n_h;\quad
\d_h={Z_h\over Z_{h-1}}(a_h-z_h);\quad U_h=({Z_h\over Z_{h-1}})^2 u_h.
\end{equation}

We perform the integration
\begin{equation}\label{ml}
\int \tilde P_{Z_{h-1}}(d\psi^{(h)}) \, e^{-\hat\VV^{(h)}
(\sqrt{Z_{h-1}}\psi^{(\le h)})}
= e^{-\VV^{(h-1)}(\sqrt{Z_{h-1}}\psi^{(\le h-1)})},
\end{equation}
where $\VV^{(h-1)}$ has the same form as $\VV^{(h)}$ and the procedure
can be iterated, as the insertion of (\ref{ml}) in (\ref{ml1})
gives an expression like (\ref{3.22})
with $h-1$ replacing $h$. The above procedure is iterated untill
a scale $h^*$ defined as the minimum $h$ such that $\g^{h}>|\s_h|$ is reached.
Then we will integrate directly the field $\psi^{(<h^*)}=\sum_{k=-\io}^{h^*}
\psi^{(k)}$ without splitting the corresponding
integration in scales (as was done
for $h>h^*$). This can be done as $g^{\le h^*}(\xx-\yy)$
verifies the bound eq(\ref{klo3}) with $h^*$ replacing $h$ \ie
it verifies the bound valid for a single scale; 
the reason is thatfor momenta larger than $O(\g^{h^*})$ the theory it is essentially a massless
theory and for momenta smaller is a massive theory with mass $O(\g^{h^*})$.
We will call {\it running coupling
constants} $\vec v_h=(U_h,\d_h,\nu_h)$ 
and {\it renormalization constants} $Z_h,\s_h$; their behaviour
as a function of $h$ can be found by an iterative equation
called {\it beta function}. Note that the irrelevant
terms are {\it not} neglected, contrary
to what is done
in the usual RG
methods, which are only approximative
and not mathematically exact. The expansion
generated by our RG is instead {\it exact} in a mathematical
sense and nothing is neglected (see Refs. \cite{[M1]} and \cite{[GM]} for details).

\subsection{The small divisor problem}
We have considered {\it irrelevant}
the terms involving two or four fermions
in the effective potential not verifying (\ref{f1}).
Looking at (\ref{ty}) we see that each addend 
contributing to the effective potential
describes the
interactions of $2n$ fermions whose momenta
{\it measured from the Fermi surface} verify
\begin{equation}\label{km}
\sum_{i=1}^{2 n}\s_i k'_i=\sum_i\s_i \o_i p_F+2 m p_F.
\end{equation}
Then (\ref{f1}) says simply that the non irrelevant terms
are only the ones in which the sum of the momenta measured
from the Fermi surface is vanishing modulo $2\pi$.
This condition seems very
natural in the {\it commensurate} case
\ie when $p_F=\pi{P\over Q}$; in that case, for $n=1,2$
if the r.h.s. of (\ref{f1})
is non vanishing modulo $2\pi$, then
it is greater than $O({1\over Q})$ so for $Q$ not too big
at least one fermion involved has a momentum far
enough from the Fermi surface.
However
things are not so simple in {\it incommensurate} case; in such
case for $n=1,2$ the r.h.s. of (\ref{ty})
can be very small (modulo $2\pi$)
for very large $m$; in other words there are terms
in the effective potential which are dimensionally relevant or marginal
involving fermions with momenta arbitrarily close to
the Fermi surface and not verifying (\ref{f1}); for instance 
$\psi^+_{k+p_F,1}\psi^-_{k-p_F+2m p_F,-1}$
with $2 m p_F+ 2 k\pi\simeq 0$ for a suitable $k$.
Then in the incommensurate case it is not clear if 
the terms not verifying (\ref{f1})
are really irrelevant (this problem is often not seen 
in literature, see for intance Ref.\cite{[VMG]}) . 
This problem, with an interely different language,
is well known in classical or celestial mechanics as the {\it small divisor
problem}, for instance in the KAM or Lindstedt 
series for invariant tori of an hamiltonian system
close to an integrable one.
It is possible to write such classical series
in terms of Feynman graphs so that the similarity becomes
very clear, see Refs.\cite{[E]} and \cite{[G0]}; the crucial difference
is that such graphs have no loops, contrary to what happens here.
Another remarkable case in which small divisors appear is in the study of
Schroedinger equation with a quasi periodic potential (very
related to our problem in the $U=0$ case);
if a
{\it Diophantine condition} is assumed on the period there are
quasi-Bloch states if $\l$ is small, see Ref.\cite{[DS]}, for
suitable values of the quasi momentum, while the
eigenstates are localized ({\it Anderson localization}) for large $\l$,
see Ref.\cite
{[PF]}. 

The fact that the contributions to the effective potential not verifying
(\ref{f1}) are {\it irrelevant} in a RG sense
means that the perturbative series
as function of 
$\l_k,\d_k,\nu_k$ and ${Z_k\over Z_{k-1}}$ and ${\s_k\over\s_{k-1}}$
are convergent in a neighborhood of the origin.
To give a complete mathematical proof of the above
statement is not straightforward, as
one has to use determinant bounds for the fermionic
truncated expectation;
one cannot simply prove that each Feynman graph admits
a finite bound as the number of Feynman graphs at order
$n$ is $O(n!^2)$, so we refer to Ref.\cite{[M1]}. However
the key idea why the terms not verifying
(\ref{f1}) are irrelevant can be understood from
an analysis based on Feynman graphs.

Each $W^{(h)}_{2n,m}$ 
admits an expansion in terms of {\it Feynman diagrams}
defined in the following way.
A $k$-th order
diagram contributing to $W^{(h)}_{2n,m}$ can be obtained from
$k$ graph-elements representing the addenda in (\ref{vbv})
or in $\RR \VV^{(0)}$ (\ref{v})
by pairing the half lines (bearing indices $h,\o,\kk'$).
The unpaired lines are called {\it external lines},
and to each paired line we associate a propagator
$g^{h_i}_{\oo_i,\oo'_i}(\kk'_i)$ (\ref{pr11}); integrating the product of these
factors over all the momenta $\kk'_i$ of the paired lines we obtain the
{\it value} of the graph, if the
expression is multiplied by a suitable sign to take into account the
Fermi statistic.
A maximal connected subset
of lines with scales $\geq h_v$
is called {\it cluster} with scale $h_v$, and
denoted by $v$. An inclusion relation can be established between the
clusters, in such a way that the innermost clusters are the clusters
with the higher scale, and so on; 
see the picture for an example of
graphs with its clusters, pictured as boxes including the paired
lines.  The half-lines (contracted or not contracted) 
are emerging by the {\it end-points},
associated to $\vec v_h$ or to the kernels of $\RR\VV^0$; if
to an end-point is associated $\vec v_k$ , the minimal cluster containing it
has scale $k$.
\def\8{\write13}
\catcode`\%=12\catcode`\{=12\catcode`\}=12
\catcode`\<=1\catcode`\>=2
\openout13=figa.ps
\8<gsave>
\8<% x1 y1 x2 y2 cambio_coordinate>
\8</cambio_coordinate{ /y2 exch def /x2 exch def /y1 exch def /x1 exch def>
\8</dx x2 x1 sub def  /dy y2 y1 sub def>
\8<x1 y1 translate dy dx atan rotate>
\8<dx 2 exp dy 2 exp add sqrt>
\8<} def>
\8<>
\8<% lx ly n normonda>
\8</normonda { /n exch def /ly exch def /lx exch def>
\8</maxang 360 n mul def  /imax 18 n mul def>
\8</fx0 [ 0 20 maxang { } for ] def>
\8</fy0 [ 0 1 imax { fx0 exch get sin } for ] def>
\8</fx [  0 1 imax { fx0 exch get maxang div lx mul } for ] def>
\8</fy [  0 1 imax { fy0 exch get ly mul } for ] def>
\8</nx fx length 1 sub def>
\8<fx 0 get fy 0 get moveto>
\8<1 3 nx {>
\8<dup dup 1 add exch 2 add 3 1 roll exch>
\8<dup fx exch get 4 1 roll fy exch get 3 1 roll>
\8<dup fx exch get 3 1 roll fy exch get exch>
\8<dup fx exch get exch fy exch get curveto>
\8<} for>
\8<} def>
\8<>
\8<% x1 y1 x2 y2 onda>
\8</onda { gsave>
\8<cambio_coordinate % st: lx>
\8<5 4 normonda stroke grestore } def>
\8<>
\8</freccia { gsave % uso: x1 y1 x2 y2 freccia>
\8<cambio_coordinate % st: l>
\8<dup 0 0 moveto 0 lineto % st: l>
\8<2 div 0 translate>
\8<15 rotate 0 0 moveto -5 0 lineto -30 rotate 0 0 moveto -5 0 lineto>
\8<stroke grestore } def>
\8<>
\8</frecciafin { gsave % uso: x1 y1 x2 y2 freccia>
\8<cambio_coordinate % st: l>
\8<dup 0 0 moveto 0 lineto % st: l>
\8<0 translate>
\8<15 rotate 0 0 moveto -5 0 lineto -30 rotate 0 0 moveto -5 0 lineto>
\8<stroke grestore } def>
\8<>
\8</punto { gsave  % uso: x1 y1 r punto>
\8<0 360 newpath arc fill stroke grestore} def>
\8<>
\8</cerchio { gsave % uso: x1 y1 r cerchio>
\8<0 360 newpath arc stroke grestore} def>
\8<>
\8</tlinea { gsave % uso: x1 y1 x2 y2 tlinea>
\8<moveto [4 4] 2 setdash lineto stroke grestore} def>
\8<>
\8</normarco { gsave newpath % l normarco>
\8<dup dup 2 div exch  3 sqrt -2 div mul % l x0 y0>
\8<3 copy 3 -1 roll add % l x0 y0 x0 y1>
\8<5 2 roll 3 -1 roll 60 120 arc % disegna l'arco; st: x0 y1>
\8<translate 15 rotate 0 0 moveto -5 0 lineto >
\8<-30 rotate 0 0 moveto -5 0 lineto>
\8<stroke grestore>
\8<} def>
\8<>
\8</cfreccia { gsave % uso: x1 y1 x2 y2 cfreccia>
\8<cambio_coordinate % st: l>
\8<normarco grestore } def>
\8<>
\8</cfrecciaspe { gsave % uso: x1 y1 x2 y2 cfrecciaspe>
\8<cambio_coordinate 1 -1 scale % st: l>
\8<normarco grestore } def>
\8<>
\8</normarcofin { gsave newpath % uso: l normarcofin>
\8<dup dup dup 2 div exch  3 sqrt -2 div mul % l l x0 y0>
\8<3 -1 roll 60 120 arc % disegna l'arco; st: l>
\8<0 translate -15 rotate 0 0 moveto -5 0 lineto -30 rotate 0 0 moveto -5 0>
\8<lineto stroke grestore>
\8<} def>
\8<>
\8<>
\8</cfrecciafin { gsave % uso: x1 y1 x2 y2 cfrecciafin>
\8<cambio_coordinate % st: l>
\8<normarcofin grestore } def>
\8<>
\8</normzigzag { % st: l>
\8<100 div dup scale 0 0 moveto>
\8<4 { 6.25 7.5 rlineto 12.5 -15 rlineto 6.25 7.5 rlineto} repeat } def>
\8<>
\8</zigzag { gsave % uso: x1 y1 x2 y2 zigzag>
\8<cambio_coordinate % st: l>
\8<normzigzag stroke grestore } def>
\8<>
\8</slinea { gsave % uso: x1 y1 x2 y2 n slinea>
\8<setlinewidth 4 2 roll moveto lineto stroke grestore} def>
\8<>
\8</sfreccia { gsave % uso: x1 y1 x2 y2 n sfreccia>
\8<setlinewidth freccia grestore} def>
\8<>
\8</semicerchio { gsave % uso: x1 y1 r semicerchio>
\8<0 180 newpath arc stroke grestore} def>
\8<>
\8</puntino { gsave  % uso: x1 y1 puntino>
\8<0.1 0 360 newpath arc fill stroke grestore} def>
\8<>
\8</mazza {gsave  % uso: l ang mazza>
\8<currentpoint translate rotate 0 0 moveto dup 0 lineto>
\8<stroke newpath 5 add 0 5 0 360 arc stroke grestore} def>
\8<>
\8</gmazza {gsave  % uso: l ang gmazza>
\8<currentpoint translate rotate 0 0 moveto dup 0 lineto>
\8<stroke newpath 10 add 0 10 0 360 arc stroke grestore} def>
\8<>
\8</ovale {gsave  % uso: r ang x y ovale>
\8<translate rotate 1 .75 scale 0 0 3 2 roll 0 360 newpath arc stroke>
\8<grestore} def>
\8<>
\8</ovales {gsave  % uso: r ang x y ovale>
\8<translate rotate 1 .5 scale 0 0 3 2 roll 0 360 newpath arc stroke>
\8<grestore} def>
\8<>
\8</arco {gsave % uso: x1 y1 x2 y2 arco>
\8<cambio_coordinate % st: l>
\8<newpath dup dup 2 div exch  3 sqrt -2 div mul % l x0 y0>
\8<3 -1 roll 60 120 arc stroke grestore} def>
\8<>
\8<1 1 scale>
\8<50 20 translate>
\8<>
\8<1 setlinewidth>
\8<20 20 50 32 cfreccia>
\8<90 32 120 20 cfreccia>
\8<50 32 90 32 freccia>
\8<120 20 20 20 cfreccia>
\8<20 20 70 120 cfreccia>
\8<120 20 70 120 cfrecciaspe>
\8<90 32 50 32 cfreccia>
\8<50 32 90 32 cfreccia>
\8<55 0 70 24 ovales>
\8<24 0 70 32 ovales >
\8<0.5 setlinewidth>
\8<70 55 73 0 360 arc>
\8<stroke>
\8<1 setlinewidth>
\8<-8 20 20 20 freccia>
\8<148 20 120 20 freccia>
\8<70 120 45 145 freccia>
\8<70 120 95 145 freccia>
\8<>
\8</Palatino-Italic findfont 9 scalefont setfont>
\8<130 100 moveto (v) show>
\8<70 53 moveto (v) show>
\8<70 12 moveto (v') show>
\8</Times-Roman findfont 7 scalefont setfont>
\8<135 97 moveto (0) show>
\8<grestore>
\closeout13
\catcode`\%=14\catcode`\{=1
\catcode`\}=2\catcode`\<=12\catcode`\>=12
\par
\hbox{\vbox to 170truept{\vfil \includegraphics{figa.ps}
}\hfill}
To each Feynman graph it is associated by the above rule
a {\it value}, and so to each subdiagram associated
to each cluster; moreover, to the lines external 
to a cluster is associated a momentum $\kk'$
smaller in modulus than one flowing in the lines internal to a cluster.
We denote by $v$
each cluster, $h_v$ its scale (\ie all the lines internal to the
cluster $v$ have scale $\le h_v$ and at least one has scale $h_v$, and the external
are larger), by $P_v$ the indices of the external lines,
by $|P_v|$ their number and
by $N_v p_F$ the sum of the momenta $\kk$ of the external
lines.
If $v'$ is the minimal cluster enclosing the cluster $v$,
the $\RR\not=1$ operation produces one or more $\kk'$
associated with the external lines (which are bounded by $\g^{h_{v'}}$)
and one or more derivatives on the propagators internal to the cluster
(giving one or more extra $\g^{-h_v}$).
At the end (see for instance Ref.\cite{[GM]})
one gets the following bound
(essentially found by power counting using
(\ref{klo3}) for bounding the propagators) 
for a graph with $k$ vertices and $2n$ external lines
with value $G^{(h)}_{2n,k}$ 
\begin{equation}\label{rom}
|G^{(h)}_{2n,k}|\le C^k\e^{k}\g^{(2-{|2n|\over 2})h}
\prod_i |\hat\f_{n_i}|\prod_{v}\g^{-(-2+{|P_v|\over
2}+z_v)(h_v-h_{v'})},
\end{equation}
where $z_v=1$ if $|P_v|=4$ and $N_v=0$; $z_v=1$ if $|P_v|=2$ and
$N_v=\o_2$, $\o_1=-\o_2$; $z_v=2$ if $|P_v|=2$, $N_v=0$; 
$z_v=0$ in all the remaining cases. The 
factor $z_v$ is due to the action on $\RR$ on each cluster 
$v$. In order to sum over all the possible
assignments of labels $h_v$ we need that 
$-2+{|P_v|\over
2}+z_v>0$ and this is true for all the clusters with two or four external lines
not
verifying (\ref{f1}) (except when $|P_v|=2$ and $\o_1=-\o_2$
in which case it is zero).
We have to improve the bound in the cases in which $-2+{|P_v|\over
2}+z_v\le 0$.
The first case we can consider is $|P_v|=2$
and $\o_1=-\o_2$, $N_v=\o_2$. 
A similar cluster can be produced only if a)there is a 
non diagonal propagator
$g^{(k)}_{\o,-\o}$ internal to the cluster $v$; b)if there is one 
or more points associated to $\hat\f_n$ internal to $v$.
In both cases this means that there is in the bounds an extra
factor
${\s_k\g^{-k}}$ for some scale $k>h_v$, see (\ref{klo4}),
and 
\begin{equation}\label{pal}
{|\s_k|\g^{-k}}={|\s_k|\over |\s_{h}|}\g^{h-k}
{|\s_{h}|\over\g^{h}}
\le {|\s_{h}|\over\g^{h}} \g^{{1\over 2}(h-k)}
\end{equation}
and the factor $\g^{{1\over 2}(h-k)}$ allows us to sum over $h_v$.

It remains to discuss the clusters with $|P_v|=2,4$
not verifying (\ref{f1});
here is where the Diophantine condition comes in.
Given a cluster $v$,  if $\sum_i \s_i\o_i p_F+2 N_v p_F\not=0$ mod. $2\pi$
then
\begin{equation}\label{dd}
|N_v|\ge C[{\g^{-{h_{v'}\over\t}}\over |P_v|^{1\over\t}}-|P_v|].
\end{equation}
In fact, by the compact support properties of the
propagators and the {\it Diophantine condition}
\begin{equation}
a_0\g^{h_{v'}}\ge ||\sum_{i=1}^{|P_v|}\s_i k'_i||_T\ge 
||2 N_v p_F+\sum_{i=1}^{|P_v|}\s_i\o_i p_F
||_T\ge C_0 (2 |P_v|+|N_v|)^{-\t}.
\end{equation}

The meaning of (\ref{dd}) is quite clear; remembering that
the momenta of the external lines is $O(\g^{h_{v'}})$,
(\ref{dd}) says that the external fields can have momenta very
close to the Fermi surface only if $N_v$ is very large.
The corresponding contribution is then very small, for the
exponential decay properties of $\hat\f_n$.
We can define a depth $D_v$ defined in the following way:
if $v$ does not contain any cluster $D_v=1$ otherwise
$D_v=1+max_{v''} D_{v''}$ where $v''$ are the clusters contained in $v$.
It is easy to see that
\begin{equation}
\prod_i|\hat\f_{n_i}|\le e^{-k |m|\over 4}
\prod_i e^{-k |n_i|\over 4}\prod_v e^{-{k|N_v|\over 2^{D_v+1}}}.
\end{equation}
Using (\ref{dd}) and the fact that $D_v\le -h_{v'}+2$ we get
\begin{equation}
\prod_i|\hat\f_{n_i}|\le e^{-k |m|\over 4}
\prod_i e^{-k |n_i|\over 4}\prod^*_v e^{-{k\g^{-h_{v'}\over\t}
\over 2^{-h_{v'}+3}}},
\end{equation}
where $\prod^*_v$ is restricted to clusters with two or four
external lines with $N_{v}\not=0$; choosing $\g^{1\over\t}2^{-1}>1$
we can associated to each of this clusters a factor  
\begin{equation}
\exp[-k\g^{-h_{v'}\over\t}2^{h_{v'}+3}]<\g^{3 h_{v'}}\le\g^{3(h_{v'}-h_v)}
\end{equation}
and this extra factor allows us to sum over the scale indices. 

In the above discussion of small divisor problem we have used the exponential decay
of $\hat\f_n$ 
but such condition could be probably
relaxed to a power law decay \ie $|\hat\f_n|\le {C_N\over |n|^N}$
for some integer $N$ \ie relaxing the analyticity condition for $\f_x$
to a differentiability one.
This is quite reasonable, by analogy with KAM theorem
which is valid not only for analytic but also for
differentiable perturbations, see Ref.\cite{[Mo]}, but the analysis would be probably much more involved.

\subsection{The flow of the running coupling constants}

The equations for the running coupling constants
are,
for $h\ge h^*$ 
\begin{eqnarray}\label{ll1}
&&\nu_{h-1}=\g\nu_h+G^{h}_\nu\quad U_{h-1}=U_h+G^{h}_U\nn\\
&&\sigma_{h-1}=\sigma_h
+G_{\sigma}^{h}\quad
\d_{h-1}=\d_h+G^{h}_\d\nn\\
&&{Z_{h-1}\over Z_h}=1+G_{z}^{h}
\end{eqnarray}
It is convenient to split $G_i^{(h)}$,
with $i=\mu,\s,\nu$ into
\begin{equation}
G^{h}_i(\mu_h,\nu_h,\s_h;...;
\mu_0,\nu_0,\s_0)=G^{1,h}_i(\mu_h,\nu_h;...;
\mu_0,\nu_0)+
G^{2,h}_i(\mu_h,\nu_h,\s_h;...;
\mu_0,\nu_0,\s_0)
\end{equation}
where we have spitted $g^{(h)}_{\o,\o}$ as in (\ref{klo1})
and 
$G^{1,h}_i$ contains no non-diagonal
propagators and only the part $g^{(h)}_{L;\o,\o}$
of the diagonal propagators; moreover,
there are no vertices with $m\not=0$; in 
$G^{2,h}_i$ are all the remaining contributions.
It is easy to check that for $i=\mu,\nu$, for $max_{k\ge h}|\vec v_k|\le\e$,
\begin{equation}\label{111}
|G^{2,h}_i|\le C[{\s_h\over\g^h}]^2\e^2.
\end{equation}
This follows from the bound (\ref{klo3}) for $C_2^h$ and from the fact
that $\nu,\mu$ are momentum conserving terms.
For $i=\s$, by symmetry reasons, 
$G^{1,h}_i\equiv 0$ and 
\begin{equation}
|G^{2,h}_\s(\mu_h,\nu_h,\s_h;...;
\mu_0,\nu_0,\s_0)|\le C |U_h \s_h|.
\end{equation}
We decompose, for $i=\mu,\nu$, 
\begin{equation}\label{ca13}
G^{1,h}_i(\mu_h,\nu_h;...;\mu_0,\nu_0)=
\bar G^{1,h}_i(\mu_h;...;\mu_0)+\hat G^{1,h}_i(\mu_h,\nu_h;...;\mu_0,\nu_0)
\end{equation}
where the first term in the r.h.s. of (\ref{ca13}) is obtained putting
$\nu_k=0$, $k\ge h$ in the l.h.s.
It is easy to see, from the fact that $g^{(h)}_{L,\o}(\xx;\yy)$
can be divided in a even part plus a correction smaller
than a factor $\g^{h\over 4}$, that
for $max_{k\ge h}|\vec v_k|\le\e$
\begin{equation}\label{78}
|\bar G^{1,h}_\nu(\mu_h;...;\mu_0)\le C\e\g^{h\over 4}.
\end{equation}
On the other hand
\begin{equation}
|\bar G^{1,h}_\mu(\mu_h;...;\mu_h)|\le C 
\e^2\g^{{h\over 2}},
\end{equation}
as one can prove using the exact solution
of the Luttinger model, see Refs.\cite{[BGPS]},\cite{[BM1]},and \cite{[GS]}.
Moreover we have that, for $i=\nu,\mu$
\begin{equation}
|\hat G^{1,h}_i(\mu_h,\nu_h;...;\mu_0,\nu_0)|\le
C\n_h |U_h|^2
\end{equation}
Finally by
a second order computation one obtains
\begin{equation}
G^{1,h}_{\s}=\s_h U_h [\b_1+\bar G^{1,h}_\s],\qquad G^{1,h}_z=U^2_h[\b_2+\bar G^{1,h}_z],
\end{equation}
with $\b_1,\b_2$ non vanishing positive constants
and $|\bar G^{1,h}_\s|\le C |U_h|$,
and
$|\bar G^{1,h}_z|\le C |U_h|$.

By using the above properties we can control the flow
of the running coupling constants. In fact, 
if 
$|\nu_k|\le C \e[{\g^{k\over 4}+{|\s_k|\over\g^k}}]
$ for any $k\ge h^*$ (which will be proved
in the following section), it follows that
there exist positive constants
$c_1,c_2,c_3,c_4,C$
such that, if $\l,u$ are small enough and $h\ge h^*$:
\begin{equation}\label{ed}
|U_{h-1}-U|<C U^{3/2};\qquad
e^{-U\b_1 c_3 h}< {|\sigma_{h-1}|\over
|\sigma_0|}< e^{-U\b_1 c_4 h};\qquad
e^{-\b_2 c_1 U^2 h} < Z_{h-1}< e^{-\b_2 c_2 U^2 h}.\nn
\end{equation}

\subsection{Determination of the counterterm $\nu$}

We show that it is possible to fix $\nu$ to a $\l$-independent
value; more exactly we show that it is possible to choose
$\nu$ as in the $\l=0$ case so that $\nu_k$ is small for any 
$k\ge h^*$. In the $\l=0$ case, $\s_k=0$
and there are no contribution to the effective potential
with $m\not=0$; calling 
$\tilde\nu_k,\tilde\mu_k$ the analogous of $\nu_k,\mu_k$, 
we can write
\begin{equation}\label{n1}
\tilde\nu_h=\g^{-h+1}[\nu+\sum_{k=h+1}^1\g^{k-2}G^{1,k}_\nu(\tilde\nu,\tilde\mu)],
\end{equation}
where $G^{1,k}_\nu(\tilde\nu,\tilde\mu)=G^{1,k}_\nu(\tilde\nu_{k},
\tilde\mu_{k};....;\tilde\nu_{0},
\tilde\mu_{0})$.
We choose
\begin{equation}
\label{n2}
\nu=-\sum_{k=-\io}^1\g^{k-2}G^{1,k}_\nu(\tilde\nu,\tilde\mu)
\end{equation}
then
\begin{equation}
\tilde\nu_h=-\g^{-h}\sum_{k=-\io}^h \g^{k-1} G^{1,k}_\nu(\tilde\nu,\tilde\mu)
\end{equation}
and  
$|\tilde\nu_h|\le C\e \g^{{h\over 4}}$,
as by (\ref{78}) 
\begin{equation}
\g^{-h}\sum_{k=-\io}^h \g^{k-1}|G^{1,k}_\nu(\tilde\nu,\tilde\mu)|\le C'
\e \g^{-h}\sum_{k=-\io}^h \g^{k}\g^{{1\over 4} k}
\le C\e\g^{h\over 4}.
\end{equation}
For the model with $\l\not=0$, for $h\ge h^*$
\begin{equation}\label{n11}\nn
\nu_h=\g^{-h+1}[\nu+\sum_{k=h+1}^1
\g^{k-2}[G^{2,k}_\nu(\nu,\mu,\s)+G^{1,k}_\nu(\nu,\mu)]]
\end{equation}
and, inserting $\nu$ given by (\ref{n2}),
\begin{equation}\label{n111}
\nu_h-\tilde\nu_h=\g^{-h+1}\{\sum_{k=h+1}^1
\g^{k-2}G^{2,k}_\nu(\nu,\mu,\s)
+\sum_{k=h+1}^1\g^{k-2}[G^{1,k}_\nu(\nu,\mu)-  
G^{1,k}_\nu(\tilde\nu,\tilde\mu)]\}
\end{equation}
We prove that, for $h\ge h^*$, 
\begin{equation}\label{ind}
|\nu_h-\tilde\nu_h|\le \e \bar C ({\s_h\over\g^h})^2,
\qquad|\tilde\mu_h-\mu_h|\le \e ({\s_h\over\g^h})^2.
\end{equation}
The proof is done by induction, assuming that it
holds for scales $\ge h+1$ and proving
by eq(\ref{n111}) that it holds for scale $h$. Looking
at the first sum in (\ref{n111}) and using (\ref{111}),(\ref{ed})
\begin{eqnarray}
&&\g^{-h}\sum_{k=h+1}^1
\g^{k-2}|G^{2,k}_\nu(\nu,\mu)|\le C_1 
\g^{-h}\e^2 \sum_{k=h+1}^1 \g^{k-2} ({\s_k\over\g^k})^2\nn\\
&&\le({\s_h\over\g^h})^2 C_2
\e^2 \sum_{k=h+1}^1 \g^{h-k} ({\s_k\over\s_h})^2\le C_3 \e^2
({\s_h\over\g^h})^2.
\end{eqnarray}

Finally we can write
$G^{1,k}_\nu(\nu,\mu)-  
G^{1,k}_\nu(\tilde\nu,\tilde\mu)=\sum_{\bar k>k} D_{\bar k,k},$
with
\begin{equation}
D_{\bar k,k}=G^{1,k}_\nu(\nu_k,\mu_k;...;\nu_{\bar k},\mu_{\bar k};
\tilde\nu_{\bar k+1},\tilde\mu_{\bar k+1};...;\tilde\nu_0,\tilde\mu_0)-
G^{1,k}_\nu(\nu_k,\mu_k;...;\tilde \nu_{\bar k},\tilde\mu_{\bar k};
\tilde\nu_{\bar k+1},\tilde\mu_{\bar k+1};...;\tilde\nu_0,\tilde \mu_0)
\end{equation}
and, by the inductive hypothesis and (\ref{rom})
\begin{equation}
\sum_{\bar k\ge k}|D_{\bar k,k}|\le C_1\bar C \e^2 \sum_{\bar k\ge k}\g^{{1\over2}(k-\bar k)}
({\s_{\bar k}
\over\g^{\bar k}})^2\le C_2\bar C \e^2 ({\s_{k}
\over\g^{k}})^2,
\end{equation}
so that the last sum in eq(\ref{n111}) is bounded by
$\bar C\e({\s_{h}
\over\g^{h}})^2$
with $\bar C=4C$,
and (\ref{ind}) is proved; from that equation
it follows that for $h\ge h^*$, then $|\nu_h|\le C\e$, so that it is possible
to have the flow of $\nu_h$ is bounded choosing a $\l$ independent
$\nu$.

\subsection{Bounds for the density}

An expansion similar to the one we have seen for the effective
potential can be defined for the Schwinger function
and hence for $\hat\r_n$. One obtains
\begin{equation}\label{x11}
\hat\r_1=\sum_{h=h^*}^1 
\int d\kk \hat g^{(h)}_{1,-1}(\kk)+
\sum_{k=1}^\io\sum_{h=h^*}^1 \hat\r^{k,(h)}_{1,-1}
\end{equation}
with $\hat\r^{k,(h)}_{1,-1}$ being sum over 
Feynman graphs similar to the one for the effective
potential in which $h$ is the lowest scale of the propagators
and $k$ the number of points.
There are two kinds of contributions.
The ones in which there are no points 
associated to $\l\hat\f_n$, which means
that there is
at least a non diagonal
propagator; such terms are bounded by $C^k\e^k|\s_h|Z_h^{-1}$. 
The remaining ones (in which there
is at least an irrelevant point) are bounded by $C^k\e^k\g^{h\over 4}$,
the factor $\g^{h\over 4}$ coming from (\ref{rom}). 

In the same way for $n\not=1$ $\hat\r_n=\sum_{k=1}^\io\sum_{h=h^*}^1
\hat\r^{k,(h)}_{n}$.
To the expansion of $\hat\r_n$ with $|n|>1$ are contributing
only graphs with at least one irrelevant point 
$\l\hat\f_{n'}$;
in fact it is not possible, by conservation of momenta,
to get a contribution to $\hat\r_n$ with $|n|>1$ 
from graphs containing only $U,\d,\nu$ vertices (taking
into account that there are non diagonal propagators, the difference
of the external momenta of such graphs can be at most $0,\pm 2 p_F$).
Hence $|\r_n^{k(h)}|\le e^{-\k/2|n|}|\s| C^k\e^k\g^{h\over 4}$.

This concludes the construction of a well defined
algorithm for computing the l.h.s. of (5) with any prefixed
precision, if $\l,U$ are small enough; in the next section we try to find a solution
$\f\in\O$ of (5) or (6) by a contraction method.

\section{Solution of the self consistence equation}
\subsection{Contraction mapping}

We have seen in \S 2.E that $\hat\r_n=\d_{n,\pm 1}
\hat\r_n^{(0)}+\sum_{k=0}^\io \hat\r^{(k)}_n$,
where $\hat\r_n^{(0)}$ is the first addend in 
(\ref{x11}) and 
$\hat\r^{(1)}_n$ is given by graphs with one point
associated to an irrelevat term $\l\hat\f_{n'}$.
We keep in the r.h.s.
of (\ref{2.5}), instead of the full expansion, just the terms with $k=0,1$,
forgetting the others for the moment.

We find natural, in order to find a solution of (\ref{2.5a}), to proceed in the
following way, calling $\F_n=\l\hat\f_n$
for $|n|>1$:

1)We consider $\s\equiv\l\hat\f_1$ as a parameter and we study
$\F_n=\l^2\hat\r^{(1)}_n(\l,U,\s,\F)$,
looking for a solution $\F(\l,U,\s)$
in the class $\O$.

2)If a non trivial solution $\F(\l,U,\s)$
is found, we 
insert it in (6) with $n=1$
looking for a solution $\s$ function of $\l,U$.

According to the above strategy,
we have to solve, for $|n|>1$
\begin{equation}\label{bn}
\hat\f_n=-\l^2 c_n^{(1)}(\s)\hat\f_n+\l\hat\r^{(1)}_n(\l,U,\s,\F),
\end{equation}
where
\begin{eqnarray}\label{bw}
&&c_n^{(1)}(\s) =\int d\kk
\Big\{
\tilde g_{1,1}^{(1)}(\kk')\,\tilde
g_{1,1}^{(1)}(\kk'+2n\pp_F)+ \sum_{\o=\pm 1} \Big[
\tilde g_{1,1}^{(1)}(\kk')\,\sum_{h=h^*}^0
{1\over Z_h}\tilde g_{\o,\o}^{(h)}(\kk'+2n\pp_F+(1-\o)\pp_F)\\
&&+  \sum_{h=h^*}^0
{1\over Z_h}\tilde g_{\o,\o}^{(h)}(\kk')\,
\tilde g_{1,1}^{(1)}(\kk'+2n\pp_F-(1-\o)\pp_F)+ \sum_{h=h^*}^0
{1\over Z_h}\tilde g_{\o,-\o}^{(h)}(\kk')\,\sum_{h'=h^*}^0
{1\over Z_{h'}}\tilde g_{-\o,\o}^{(h')}(\kk'+2n\pp_F)\nn\\
&&+ \sum_{h=h^*}^0
{1\over Z_h}
\tilde g_{\o,\o}^{(h')}(\kk')\,\sum_{h'=h^*}^0
{1\over Z_{h'}}\tilde g_{\o,\o}^{(h')}(\kk'+2n\pp_F) +
\sum_{h=h^*}^0
{1\over Z_h}\tilde g_{\o,\o}^{(h)}(\kk')\,
\sum_{h'=h^*}^0
{1\over Z_{h'}}\tilde g_{-\o,-\o}^{(h')}(\kk'+(2n+2\o)\pp_F)
\Big]\Big\}\nn
\end{eqnarray}
and
\begin{equation}\label{bo1}
\hat\r^{(1)}_n=\sum_{h,h'=h^*}^1 {1\over Z_h Z_{h'}}
\sum_{m\not=0,n}\sum_{\o_1,\o'_1\atop\o_2,\o'_2}
\d_{2 m-\o_1+\o'_1-\o_2+\o'_2,2n}\l\hat\f_m
\int d\kk' g^{(h)}_{\o_1,\o'_1}(\kk')
g^{(h')}_{\o_2,\o'_2}(\kk'+(2m +\o'_1-\o_2)p_F),
\end{equation}
where $g^{(1)}(\kk)=g^{u.v.}(\kk)=g^{(1)}_{\o,\o}(\kk')$ if $\kk=\kk'+\o p_F$.
We shall rewrite (\ref{bn}) as
\begin{equation}
\F_n={\l^2\over 1+\l^2 c_n^{(1)}}\hat\r^{(1)}_n(\l,U,\s,\F).\label{it}
\end{equation}
and we will look for a solution $\F(\l,U,\s)$ of such equation by applying
an iterative method.

For a fixed $L$, $\F$ is a finite sequence of $L-3$ elements, which can be
thought as a vector in $R^{L-3}$.
We consider
the space $\FF=C^1(R^{L-3})$ of $C^1$-functions of 
$\hat\f_n$, $n=2,3,...,L_i-1$. 
We shall define a norm in $\FF$ 
for any $\l,\s$ different from zero
\begin{equation}
\|\F\|_\FF= \sup_{|n|>1} \left\{
e^{|n||\log|\l||\over 10}[|\s|^{-1} |\F_n(\s)|+|{\partial\F_n\over
\partial\s}|]\right\}.
\end{equation}
We shall also define
$$\BB = \{\F\in\FF : \|\F\|_\FF\le 1\}\; ;$$
The solutions of (\ref{it}) can be seen
as fixed points of the operator ${\bf T}_{\l}: \FF\to \FF$,
defined by the equation:
\begin{equation}
[{\bf T}_{\l}(\F)]_n(\s) =
{\l^2 \tilde\r_n^{(1)}(\l,U,\s,\F) \over (1+\l^2 c_n^{(1)}(\s))},
\end{equation}

In the sum over $m$ in (\ref{bo1})
$m\not=n$ so that there is at least
a non-diagonal propagator, so that by (\ref{ed}), if  
$\o'_1=-\o_1$ (say),
\begin{equation}\label{bo2}
\sum_{h,h'=h^*}^1{1\over Z_h Z_{h'}}|\int d\kk' g^{(h)}_{\o_1,-\o_1}(\kk')
g^{(h')}_{\o_2,\o'_2}(\kk'+(2m +\o'_1-\o_2)p_F)|\le C
\sum_{h=h^*}^1 [|\s_h\g^{-h}|^2+|\s_h\g^{-h}|]\le C'.
\end{equation}
Considering $c^{(1)}_n$ (\ref{bw}),
it is easy to see that the integrals in the first four
terms of (\ref{bw}) are bounded by constants, as there is
at least a non diagonal propagator or an ultraviolet one;
for the fifth integral the bound is
\begin{equation}
\sum_{h=h^*}^0
|\int d\kk'{1\over Z_h}
\tilde g_{\o,\o}^{(h')}(\kk')\,\sum_{h'=h^*}^0
{1\over Z_{h'}}\tilde g_{\o,\o}^{(h')}(\kk'+2n\pp_F)|\le
C_1 \sum_{h=h^*}^1 {1\over (Z_h)^2}\le {2 C_1\over c_1 \b_2 U^2}.
\end{equation}
A similar bound is found
also for the last integral of (\ref{bw}).

Then from the above bounds, if ${\l^2\over U^2}$
is small enough one has
\begin{equation}
{\l^2\over 1+\l^2 c_n^{(1)}}\le 2\l^2.
\end{equation}
We find a solution of (\ref{it}) as the fixed point
$\bar\F$ of ${\bf T}_\l$ in $\BB$. Note that 
$\BB$ is invariant under the action of ${\bf T}_\l(\F)$, 
as by (\ref{bo2})
\begin{equation}
|{\bf T}_{\l,n}(\F)|\le 2 C \l^2 
[\sup_{k=\pm 1,\pm 2}{|\l|^{|n+k|\over 10}\over |\l|^{|n|\over 10}}]
|\l|^{|n|\over 10}|\s|\le {1\over 2}|\l|^{|n|\over 10}|\s|.
\end{equation}
so that for $\l$ small enough
\begin{equation}
||{\bf T}_{\l}(\F)||\le 1
\end{equation}
Moreover, by a similar argument
\begin{equation}
\|{\bf T}_{\l}(\F)-{\bf T}_{\l}(\F')\|_\FF \le {1\over 2} 
\|\F-\F'\|_\FF,
\end{equation}
so that ${\bf T}_{\l}$ is a contraction.
It is also evident that ${\bf T}_{\l}(0)\in\BB$.
Hence, by the contraction mapping principle, there is
a unique fixed point $\bar\F$ of ${\bf T}_{\l}$ in $\BB$, which can
be obtained as the limit of the sequence $\F^{(k)}$ defined through the
recurrence equation $\F^{(k+1)}={\bf T}_{\l}(\F^{(k)})$, with any initial
condition $\F^{(0)}\in\BB$. If we choose $\F^{(0)}=0$, we get
\begin{equation}
\|\bar\F\|_\FF \le \sum_{i=1}^\io \|\F^{(i)}-\F^{(i-1)}\|_\FF\le \sum_{i=1}^\io {1\over 2^{i-1}} \|\F^{(1)}\|_\FF \le
\|\F^{(1)}\|_\FF,
\end{equation}
which immediately imply that $\|\bar\F\|_\FF\le 1$.
Then if ${U^2\over \l^2}$ is large enough
there is a solution $\F$ of (69)
which is a function of $\l,U,\s$; 
in the following section we insert it in (6) with $n=1$
looking for a solution $\s$ which is function of $\l,U$.

\subsection{Determination of $\l\hat\f_1$}
We have now to insert $\F(\l,U,\s)$ found above
in (6) with $n=1$, in order to find $\s$ 
in a self consistent way.
We can rewrite (\ref{2.5a}) as
\begin{equation}\label{bn1}
\hat\f_1=-\l^2 c_1^{(1)}(\s)\hat\f_1+\l\hat\r^{(1)}_1(\l,U,\s,\F),
\end{equation}
where
\begin{eqnarray}
&&c_1^{(1)}(\s) = \int d\kk' 
\Big\{ \sum_{h=h^*}^0{\tilde g_{-1,1}^{(h)}(\kk') \over \s} +
\tilde g_{1,1}^{(1)}(\kk')\,\tilde g_{1,1}^{(1)}(\kk'+2\pp_F)\\
&&+ \sum_{\o=\pm 1} \Big[
\tilde g_{1,1}^{(1)}(\kk')\,\sum_{h=h^*}^0
{1\over Z_h}\tilde g_{\o,\o}^{(h)}(\kk'+(3-\o)\pp_F) +\nn\\
&&
\sum_{h=h^*}^0{1\over Z_h}\tilde g_{\o,\o}^{(h)}(\kk')\,\tilde g_{1,1}^{(1)}(\kk'+(1+\o)\pp_F)
\Big]\Big\}\nn\\
&&= -F(\s,\l,U)+\tilde c_1(\s)\nn
\end{eqnarray}
and more explicitly
\begin{eqnarray}\label{hj}
&&F(\s,\l,U)\equiv \sum_{h=h^*}^0\int d\kk'{\tilde g_{-1,1}^{(h)}(\kk') \over \s}=\\
&&\sum_{h=h^*}^0
\int d\kk {1\over Z_h}{\s_h \over \s}{f_h(\kk)\over k_0^2+\sin^2 k'+(1-\cos
k')^2+\s_h^2},\nn
\end{eqnarray}
where $\s_h,Z_h$ verify (\ref{ed}),
and $|\tilde c_1|\le C$ where $C$ is a constant.

Then by (\ref{hj}) and (\ref{x11}) we obtain
\begin{equation}
{\l^2\over\h_1}[
({|\s|\over A})^{-\h_2}-1][a^{-1}+U f_1(\l,U,\s)]+\l\s f_2(\l,U,\s)
=1,\label{ffa1}
\end{equation}
with $|f_1|, |f_2|\le C$
and $\h_1=\b_1 U+\tilde\h_1$, $\h_2=\b_1 U+\tilde\h_2$,
$|\tilde\h_1|,|\tilde\h_2|\le C U^2$, and $C$,
$a,\b_1,A$ are positive constants.               
(\ref{ffa1}) is a non-BCS or {\it anomalous} self consistence equation
describing a superconductor whose normal state is a Luttinger liquid;
if $U=0$ it reduces to a standard BCS equation.
If the electron-electron interaction is 
larger than the electron-phonon interaction,\ie if
${|\l|\over |U|}$ is small, 
thus the
result is very sensitive to the sign of $U$. 

a) In
the {\it attractive} case $U<0$ there is {\it no solution} $|\s|<1$
if ${\l^2\over U^2}$ is small enough.
In fact all the terms in the l.h.s. of (\ref{ffa1}) can be taken arbitrary
small if $U,\l$ and ${\l^2\over |U|}$ are small enough so there is no
$|\s|\le F_0$ solving (\ref{ffa1}).
Note that this is also true considering (5),
\ie there is no solution $\f\in\O$ of (5) if 
${\l^2\over U^2}$ is small enough
for any $\f\in\O$.

b) In the {\it repulsive} case $U>0$ for ${\l^2\over |U|^2}$ small enough
there is a solution of the form (\ref{jk}), as $\b,L\to\io$.
Note that this expression for the gap 
is very different with respect to the
BCS like form of the $U=0$ case; in particular it is as large as 
$A e^{-{a\over\l^2} |{\l^2\over a\h}||\log||{\l^2\over a\h}|}|>>A
e^{-a\over\l^2}$.
A similar expression for the gap appears in the interacting
Kondo problem, see Ref.\cite{[LT]} and in superconductors whose
ground state is a Luttinger liquid, see Ref.\cite{[M2]}. It is easy to check
that corresponding to such solution the Hessian 
is positive definite.

Finally I discuss shortly how to solve the full (\ref{2.5})
without truncating the r.h.s. at the second order.
At finite $L$ we can 
use a contraction method to find a fixed point  of (\ref{2.5}).
However the fact that for the higher order terms we have only bounds,
and we cannot make use their exact expressions like was done for (\ref{2.5a}),
has the effect that we can only get a weaker bound for the decay of the solution
\ie a power law decay 
$|\l\hat\f_n|\le {C_N\s\over |n|^N}$ 
instead than an exponential one. 
This was also what we could get in Ref.\cite{[BGM2]} in the commensurate case.
In the incommensurate case this forbids to take the $L\to\io$
limit, just because we have proved in Ref.\cite{[M1]} the convergence of $\hat\r_n$ in that limit only if
$\l\hat\f_n$ have an exponential decay; on the contrary in the 
commensurate case
convergence holds with a weak decay condition and the limit can be taken.
This problem seems however merely technical, and it could
be solved proving convergence 
under power law decay (see discussion at the end of \S 2.B ); 
this would allow to prove
Peierls instability in the incommensurate case 
without truncating the density expansion.

\section{Conclusions}

In the perturbative regime of small $U$
there is Peierls instability then for $U\ge U_{c,1}$ and for $U\le U_{c,2}$
there is not. It is an interesting open question
whether $U_{c,1}=U_{c,2}=U_c$ or not, and if they are equal weather $U_c$
is greater, smaller or equal to zero.
We are at the moment not able to answer to this question; we
can only say that when $U=0$ or $U$ small with respect to
$\l^2$ the iterative method for solving (\ref{2.5a})
is not working. Let us consider
in fact the $U=0$ case. In that case (\ref{ffa1}) with $n=1$ is a BCS like 
equation and gives
$|\s|=A e^{-[a+g]\over \l^2}$,
where $|g(U,\l)|\le C|\l|$.
However we are not able to find a solution
to the equation for $\F$; the contraction method
used in the large ${U\over\l}$
case here fails as there are integers $n$ such
that $2 n p_F$ is very close to $2 p_F$ modulo $2\pi$
(of course analogous considerations
can be done if is close to $-2 p_F$)
\ie 
$||2 n p_F- 2 \o p_F||_T$ can very small. 
For such $n$, $c_n^{(1)}$ (\ref{bw}) computed
at $U=0$ (\ie $Z_h=1,\s_h=\s$)
can be written as
\begin{equation}
c_n^{(1)}=c_1^{(1)}+b+O({||2 n p_F- 2 p_F||_T\over \s})+O(\l),\label{l}
\end{equation}
where $b$ is a constant.
As $1+\l^2 c_1=O(\l^2)$ 
we have that
$${\l^2\over 1+\l^2 c_n}=O(1)$$ 
instead $O(\l^2)$ as it is in the preceding section;
so it is not clear how a contraction method could 
be applied (an explicit computation
shows that ${\l^2\over 1+\l^2 c_n}$ times the coefficient of $\hat\f_{n+2}$
tends to $1$ when $\a_n\to 0$).
This fact could be not simply a technical
problem linked to the method. 
The idea underling the analysis of the preceding section, 
which is essentially the Peierls
idea, is that the process involving the exchange of $2 p_F$
is the dominant one, and the others are corrections.
However, in the incommensurate case for very large
$n$, $2 p_F$ or $2 n p_F$ are almost the same, due to Umklapp
scattering, so it is not clear physically why the first harmonic
is the dominant one. Technically this means that there are
large $n$ for which $\hat\r_n\simeq \hat\r_1$ (see (\ref{l}) in which
$b$ is negligible with respect to $c_1$ which is log-diverging)
so that the self-consistence equation $\hat\f_n=\l\hat\r_n$
for such $n$ or for $n=1$ seems similar, and it is not clear
why $|\hat\f_n|<<|\hat\f_1|$, as it should be in order $\f_x$
to be analytic. This problem is absent in 
the {\it commensurate} case; in that case, see Ref.\cite{[BGM2]}, it is possible to
see that $|c_n^{(1)}|\le C\log Q$ so that
for $Q\le e^{-{1\over |\l|}}$ then
${\l^2\over 1+\l^2 c_n}\le |\l|$ and the contraction method 
will work, so that Peierls instability is proved. 

The mechanism why the above difficulty is avoided
in the Holstein-Hubbard model with $U>0$ and ${U\over\l^2}$
large, so that Peierls instability is proved in the incommensurate
case, is that 
the harmonic with $n=1$ has a non trivial flow,
and becomes larger, while $\hat\f_n$ for $|n|\ge 2$
has no flow; this is in a sense a consequence of the 
diophantine condition.

The difficulty in finding a solution to (\ref{2.5a}) when $U=0$
could mean that
there is Peierls instability only for $U>U_c$
with $U_c>0$, but of course a deeper analysis is necessary
to conclude. Note that
in the $d=\io$ limit of the Holstein-Hubbard model
an incommensurate CDW is found, see Ref.\cite{[FK]}, only for $U>U_c$
with $U_c$ non vanishing positive. On the other hand
the existence of Peierls instability when $U=0$ is usually 
supported
by the analogy of the Holstein model with
the {\it Frenkel-Kontorova} chain
in which the
extremality condition for the energy gives
a dynamical system known as {\it standard map}. For this model
the existence of an incommensurate phonon field minimizing 
the energy is an application of {\it KAM} theorem
in the perturbative regime (and in the strong coupling regime
of the so-called Aubry-Mather theorem, in which $\f_x$
is not smooth but it
has
infinite many discontinuities).


\section{References}

\begin{references}

\bibitem{[A]} V.I. Arnold, {\it Russ. Math. Surveys} 18,9-36 (1963)

\bibitem {[AL]} S. Aubry, P.Y. Le Daeron:
{\it Phys. D} {\bf 8}, 381--422 (1983)


\bibitem{[AAR]}S. Aubry, G. Abramovici, J. Raimbault:
{\it J. Stat. Phys.} {\bf  67}, 675--780 (1992)

\bibitem{[B]} A. Bianconi et al: Stripes and related phenomena, edited
by Bianconi and Saini, Kluwer, New York (2000)

\bibitem{[BGM1]} G.Benfatto,G.Gentile, V.Mastropietro.
{\it J. Stat. Phys.} 89, 655-708 (1997)

\bibitem{[BGM2]} G.Benfatto,G.Gentile, V.Mastropietro.
{\it J. Stat. Phys.} 92, 1071-1113 (1998)


\bibitem{[BM1]} F.Bonetto, V.Mastropietro: {\it Commun. Math. Phys.}, 172,
57-93 (1995)

\bibitem{[BM2]} F.Bonetto, V.Mastropietro: {\it Phys. Rev. B}, 56,
3, 1-43 (1996)

\bibitem{[BGPS]}G. Benfatto, G. Gallavotti, A. Procacci, B. Scoppola:
{\it  Comm. Math. Phys.} {\bf  160}, 93--171 (1994)

\bibitem{[CG]} A.H. Castro-Neto,F.Guinea: {\it Phys. Rev. Lett.}
80,18, 4040-4043 (1998)

\bibitem{[DS]} E.Dinaburg, Y.G. Sinai; {\it Funct. Anal. and its Appl.},
9,279-289 (1975)

\bibitem{[E]} L.H. Eliasson; {\it Math. Phys. El. Journ.} 2 (1998)

\bibitem{[F]} F H. Frolich: Proc. Roy. Soc., A223,296, (1955)

\bibitem{[G0]} G. Gallavotti; {\it Comm. Math. Phys.} 164, 145--156 (1994) 

\bibitem{[G]} G.Gallavotti, Elementary mechanics, Springer (1983)

\bibitem{[GM]} G.Gentile, V.Mastropietro:{\it Physics Reports}, 352, 273-437
(2001)

\bibitem{[GS]} G. Gentile, B. Scoppola:
{\it Comm. Math. Phys.} {\bf 154}, 153--179 (1993).

\bibitem{[FR]}R.H. Friend: in Solitons and condensed matter physics, Ed A.R.Bishop
and T. Schneder, Springer, (1978)

\bibitem{[FK]} J.K. Freericks, M. Jarrell cond mat/9502098 

\bibitem{[KL]}T. Kennedy, E.H. Lieb:
{\it Phys. Rev. Lett.} {\bf 59}, 1309--1312 (1987).

\bibitem{[LN]}E.H. Lieb, B. Nachtergale; {\it Phys. Rev. B},51,8,4777-4791
(1995)

\bibitem{[LT]} D.Lee, J.Toner, {\it Phys. Rev. Lett.} 69 (23) 3378--3381 (1992)

\bibitem{[LRA]}
P.A.Lee, T.M. Rice, P.W.Anderson; Solid State Comm. 14,703 (1974)

\bibitem{[M1]} V.Mastropietro Commun. Math. Phys. 201 81-115 (1999)

\bibitem{[M2]} V.Mastropietro Mod. Phys. Lett. 201 81-115 (1999)

\bibitem{[Mo]} J. Moser.  {\it Nachr.Akad.Wiss. Gottingen Math-Phys} 1-20 (1962))

\bibitem{[P]}
R.E. Peierls: in Quantum Theory of solids
, O.U.P., (1955)

\bibitem{[PCGD]}
A.Perali, C.Castellani,C. Di Castro,M.Grilli, 
{\it Phys. Rev.} B 54,16216 (1996)

\bibitem{[PF]} L.Pastur, A. Figotin: Spectra of Random and
quasi-periodic operators, Springer, Berlin (1991)


\bibitem{[T]} G.A.Toombs: Phys. Rep. 40C,181
(1978)

\bibitem{[VMG]}
J.Vidal, D.Moihanna,T. Giamarchi: cond-mat/9905080

\end{references}
\end{document}